\documentclass[12pt,a4paper]{article}
\usepackage{a4wide}
\usepackage{latexsym}
\usepackage{epsf}
\usepackage{amssymb}
\begin{document}
\def\be{\begin{equation}}
\def\ee{\end{equation}}
\def\bea{\begin{eqnarray}}
\def\eea{\end{eqnarray}}
\newcommand{\fsl}{{\hspace{-9pt}\slash}}

\def\pd{\partial}
\def\a{\alpha}
\def\b{\beta}
\def\g{\gamma}
\def\d{\delta}
\def\m{\mu}
\def\n{\nu}
\def\t{\tau}
\def\l{\lambda}
\def\O{\Omega}
\def\r{\rho}
\def\s{\sigma}
\def\e{\epsilon}
\def\scri{\mathcal{J}}
\def\cM{\mathcal{M}}
\def\tcM{\tilde{\mathcal{M}}}
\def\RR{\mathbb{R}}

\hyphenation{re-pa-ra-me-tri-za-tion}
\hyphenation{trans-for-ma-tions}


\begin{flushright}
IFT-UAM/CSIC-02-43\\
hep-th/0301123\\
\end{flushright}

\vspace{1cm}

\begin{center}

{\bf\Large Codimension Two Holography}

\vspace{.5cm}

{\bf Enrique \'Alvarez, Jorge Conde and Lorenzo Hern\'andez }

\vspace{.3cm}

\vskip 0.4cm

{\it  Instituto de F\'{\i}sica Te\'orica UAM/CSIC, C-XVI,
and  Departamento de F\'{\i}sica Te\'orica, C-XI,\\
  Universidad Aut\'onoma de Madrid 
  E-28049-Madrid, Spain }

\vskip 0.2cm

\vskip 1cm

{\bf Abstract}
A 
holographic interpretation for
some specific Ricci flat string backgrounds of the form $A_6\times C_4$ is proposed. 
The conjecture is that there is a Four-dimensional Euclidean Conformal Field Theory (ECFT) defined 
on a codimension two
{\em submanifold} of the manifold $A_6$ (where one of the two 
remaining {\em holographic} coordinates of $A_6$ is timelike, and the other one spacelike), with 
central charge proportional
 to the radius of curvature of the six-dimensional manifold, $c\sim l^4$.
\end{center}

\begin{quote}

\end{quote}


\newpage

\setcounter{page}{1}
\setcounter{footnote}{1}
\newpage
\section{Introduction}
Let us fix our attention on a  string background 
described as a Ricci flat manifold of the form
\be
M_{10}\equiv A_{6}\times C_{4}
\ee
where $C_4$ is an internal compact manifold, and $A_6$ will be denoted 
by the name {\em ambient space}. There are no Ramond-Ramond backgrounds excited,
so that this background is valid for all types of strings. In this ambient space lives a codimension two
{\em euclidean} four-manifold, which will be interpreted as the {\em spacetime} $M_4\subset A_6$.
The spacetime coordinates will be denoted by 
$x^i \equiv\vec{x}$  and its metric by $g_{ij}dx^i dx^j$;  whereas the extra two coordinates of the ambient 
space by $\rho\in\mathbb{R}^{+}$ 
and $t\in\mathbb{R}^{+}$, where $\rho$ is spacelike and $t$ timelike.
There is then a natural {\em boundary} defined in this patch by
\be
\pd A_6\equiv \{\rho=0\}
\ee 
It will be moreover assumed that the ambient space metric (cf. Appendix) can be written as:
\be\label{ambient}
ds^2=\frac{t^2}{l^2} ds^2(x,\rho) +\rho dt^2 + t d\rho dt
\ee
where the metric  induced on the hypersurfaces $\Sigma\equiv \{t=const.\}$,
namely $ds^2(x,\rho)\equiv h_{ij}(x,\rho)dx^i dx^j $ is such that it 
reduces to the (euclidean) spacetime metric
 on $\rho=0$:
\be
ds^2(x,\rho=0) = g_{ij}(x) dx^i dx^j.
\ee
The set of spaces obeying these restrictions is a non-empty set, and we provide several examples
in the appendices.
\par
The purpose of the present paper is to show, first of all, that there 
are diffeomorphisms on $A_6$ that reduce to Weyl transformations on the boundary.
This strongly suggests that there is some conformal theory associated to the said boundary.
\par
We then further discuss the boundary energy momentum tensor, and show, by discarding 
a divergent part, that it is proportional to the four-dimensional conformal anomaly for conformally 
invariant matter.
\par
We then conlude with some comments on the relationship of this approach with the usual 
AdS/CFT of Maldacena's.
\section{PBH Diffeomorphisms}
One way of understanding the fact that diffeomorphism invariants on some
 manifold give rise to conformal invariants in some other manifold which
is in a precise sense  the boundary of the former one, is by stablishing the
existence of the so-called Penrose-Brown-Henneaux (PBH) 
(\cite{penrose}\cite{brown}) diffeomorphisms; that is, 
diffeomorphisms that reduce to conformal 
transformations on the boundary. 
\par
This approach has been pioneered in a related context, namely for
the study of the bulk space in AdS/CFT in 
(\cite{imbimbo}\cite{Schwimmer}). Let us examine it in the present context.
We shall perform the computations for arbitrary $A_{n+2}$ in the sequel,
although we shall be mostly interested in the case $n=4$.

\par
The most general diffeomorphism that maintains the  coordinate gauge  
(that is, such that
$\delta g_{tt}=\delta g_{\rho t}=\delta g_{\rho\rho}=\delta g_{ti}=
\delta g_{\rho i}=0$)\footnote{We are refering here to the components of the
$(n+2)$-dimensional ambient metric, not to be confused with the zero mode at
the boundary} is generated
by the vector
\be
\xi=\left(-a(x)t+b(x)\right) \pd_{t}+ \left(
2a(x)\r-\frac{b(x)\r+c(x)}{t}\right) \pd_{\r}
+\xi^{i}(t,\r,x)\pd_{i}
\ee
where
\bea
&&\frac{2 t^2}{l^2}h_{ij}\frac{\pd}{\pd\rho}\xi^j -t^2\pd_i a+t\pd_i b=0\nonumber\\
&&\frac{2 t^2}{l^2}h_{ij}\frac{\pd}{\pd t}\xi^j +\rho\pd_i b-\pd_i c=0
\eea
and, besides, $\xi^i|_{\rho=0}=0$.
\par
This means that the induced variation on the metric $h_{ij}$  is given by:
\be
\delta h_{ij}(\rho,x)=\left( 2a(x)-\frac{b(x)}{t} \right)   h_{ij}
+\left( 2a(x)\r-\frac{b(x)\r+c(x)}{t} \right) \pd_{\r}h_{ij}
+\stackrel{
  ^{(h)}}{\nabla_{i}}\xi_{j}+\stackrel{
  ^{(h)}}{\nabla_{j}}\xi_{i}
\ee
Please note that we have defined here
\be
\frac{t'^{2}}{l^{2}}h'_{ij}(\r',x')=\frac{\pd x^{k}}{\pd x'^{i}}
\frac{\pd x^{l}}{\pd x'^{j}}\frac{t^{2}}{l^{2}}h_{kl}(\r,x) +
O(\xi^{2})
\ee

\par
If we compute the action of such a transformation on the metric at 
the boundary $\rho=0$, it results in
\be
\d g_{ij}=(2a(x)-\frac{b(x)}{t})g_{ij}
+(\frac{c(x)}{t})h^{(1)}_{ij}
+\ 2\nabla_{(i}\xi_{j)}\mid _{\r=0}
\ee
where
\bea
&&g_{ij}=h_{ij}(\rho=0)\nonumber\\
&&h^{(1)}_{ij}=\frac{d h_{ij}}{d\rho}|_{\rho=0}
\eea
and the covariant derivative is with respect to the metric $g$.
In order to
obtain a pure Weyl transformation on the boundary we have to choose 
\begin{eqnarray}
b(x)=c(x)=0\nonumber \\
\left. \xi(\r,x)\right|_{\r =0}=0
\end{eqnarray}
We have not analyzed the interesting possibility of keeping $b\neq 0$, which
leads to time-dependent Weyl transformations at the boundary.
\par

To summarize, the most general PBH is given by:
\be
\xi=\epsilon(x)[-t\frac{\pd}{\pd t}+ 2\rho\frac{\pd}{\pd \rho}]+\xi^i(\rho,x)\frac{\pd}{\pd x^i}
\ee
\par

 It is curious to notice that the boundary
 $B_{n+1}\equiv\pd A_{n+2}$ is a $n+1$-dimensional theory with Lorentzian signature,
involving the coordinates $x^i$ as well as the time $t$. The metric on this
 boundary is degenerate, namely
\be
ds^2=\frac{t^2}{l^2}h_{ij}(x,\rho)dx^i dx^j
\ee 
The timelike coordinate only appears in this metric as a multiplicative factor,
and never in the metric $h_{ij}$ itself, so that its zero mode, namely 
$g_{ij}(x)\equiv h_{ij}(x,\rho=0)$ is also {\em a fortiori} time independent.
 Besides, the conformal 
transformations do not depend on time at all. For all that matters
at the boundary, 
time is just an external parameter
\par
The essential part of the PBH diffeomorphism\footnote{That is, suppressing
 purely
spatial diffeomorphisms generated by $\xi^i \pd_i$.} however, is
\be
\xi= \epsilon(x)[-t\frac{\pd}{\pd t}+2\rho \frac{\pd}{\pd \rho}]
\ee
which mixes the two {\em holographic} coordinates $(t,\rho)$ in a particular
combination. This is the root of many properties of this construction.
\par
In conclusion, any covariantly defined theory in the ambient space generates
a Weyl invariant one on the boundary with the qualifications as above. 

\section{The Regularized Boundary}
The true $n+1$-dimensional boundary, $B_{n+1}\equiv\{\rho=0\}$ is
a null surface. The null character of the normal vector is however an 
isolated fact of the normal vector field. This suggest the
consideration of the surface $\rho=\e$, which we shall call 
the regularized boundary $B_{\e}$.
\par
The first thing to notice is that this surface is now timelike, 
with normal vector
\be
n_{\e}=\e^{-1/2}\frac{\pd}{\pd t}-\frac{2\e^{1/2}}{t}\frac{\pd}{\pd \r}
\ee
Only in the strict limit $\e=0$ the surface becomes null. This is however the
most  natural extension of the normal to a vector field, and, as we shall see,
it is essential to regularize the action in some way in order to properly
define physical quantities. Correspondingly, the induced metric is now non-degenerate, and given by:
\be
ds^2_{B_{\e}}=\e dt^2+\frac{t^2}{l^2}h_{ij}dx^i dx^j.
\ee
\par

\section{The Brown-York quasilocal energy}
There is a convenient definition of a quasilocal gravitational energy 
due to Brown and York (BY), \cite{Brown} which, although not conserved in 
general, embodies a large fraction of the asymptotic symmetries of the 
gravitational field. As will be shown in the sequel, this quantity is
precisely defined on codimension two surfaces. 
\par
This general idea has been succesfully exploited by 
Brown and Henneaux in \cite{brown} to associate a CFT to $AdS_3$, and by
Balasubramanian and Kraus \cite{Balasubramanian} to introduce a boundary 
energy-momentum tensor.
\par
This definition of energy shares with the usual Arnowitt, Deser and Misner
(ADM) definition, which is valid for asymptotically flat spacetimes, the fact
 that it is defined with respect to a foliation by a family of 
spacelike surfaces $\Sigma$ and then it is expressed as an integral over the
 boundary $\pd\Sigma$, but it differs in the detail, and can be also applied to 
non asymptotically flat situations. 
\par
In order to properly define a variational principle,
it is convenient to consider a region of the total spacetime $A_{n+2}$, say $M$,
bounded by two initial and
final spacelike hypersurfaces, $\Sigma_i$ and $\Sigma_f$, which in our case 
are $(n+1)$-dimensional, and a timelike boundary, $B_{n+1}=\pd M$. Please refer to the included figure for the 
geometrical setup.

\par
\begin{figure}[!ht] 
\begin{center} 
\leavevmode 
\epsfxsize= 5cm
\epsffile{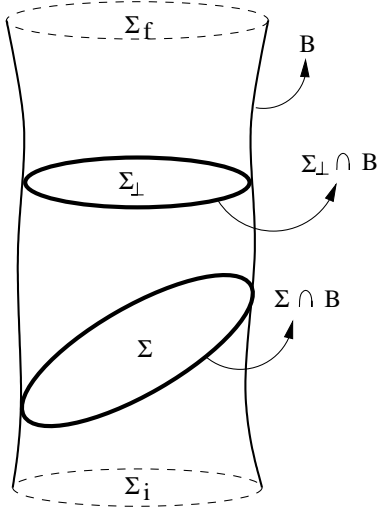} 
\caption{{\em The hypersurfaces $\Sigma_{\perp}$,  contrary to other foliations
of the spacetime $\Sigma$ (such as $t=constant $), enjoy an orthogonal
intersection with the spacetime boundary $B$}}
\label{fig:uno} 
\end{center} 
\end{figure}
\par

Actually, instead of considering the $(n+1)$-dimensional boundary $B_{n+1}\equiv\{\rho=0\}$ 
which is, as we
have seen, a null surface, we shall consider previously introduced 
{\em regularized boundary}.
The extrinsic curvature $\Theta_{ij}\equiv \frac{1}{2}\pounds_n g_{ij}$ is 
given by:
\be
\Theta_{ij}= \frac{t}{l^2 \epsilon^{1/2}}[h_{ij}-\epsilon h^{\prime}_{ij}]
\ee
where $h^{\prime}_{ij}\equiv\frac{d}{d\rho}h_{ij}$.
The corresponding boundary energy-momentum tensor, defined as
\be
\tau_{ab}\equiv\frac{1}{\kappa^2}
[\Theta_{ab}-\Theta g_{ab}]
\ee
 has got components
\bea
&&\tau_{tt}=-\frac{1}{\kappa^2}\frac{\epsilon^{3/2}h^{kl}h^{\prime}_{kl}-
n \epsilon^{1/2}}{t}\nonumber\\
&&\tau_{ij}=-\frac{1}{\kappa^2}[\frac{(n-1)t}{\epsilon^{1/2}l^2}h_{ij}+
\frac{\epsilon^{1/2}t}{l^2}(h^{\prime}_{ij}-(h^{kl}h^{\prime}_{kl})h_{ij})]
\eea
It is to be remarked that in spite of its name, this boundary 
energy-momentum tensor is a quantity that refers to a manifold 
such as the regularized boundary, of Lorentzian signature.
In order to define a energy in the BY sense , 
we still need to foliate the complete 
ambient spacetime with a family of spacelike surfaces $\Sigma$, and 
the energy so defined depends on the foliation in a nontrivial way.. 
\par
The choice $\Sigma\equiv\{t=const\}$ is not adequate, first of all, 
because these surfaces are null  and in addition because they do not enjoy 
an orthogonal 
intersection with the boundary of the spacetime . This last point, 
although technical,
 greatly complicates the analysis, and makes the definition of 
energy less useful.
\par
Both problems could be remedied  at one fell swoop if we consider instead
the surfaces $\Sigma_{\perp}$ generated by the vector.
\be
u\equiv\rho^{-1/2}\frac{\pd}{\pd t}
\ee
which are easily found to correspond to
\be
\Sigma_{\perp}\equiv \{t\rho^{1/2}=L,\rho>0\}.
\ee
(where $L$ is an arbitrary constant).
The quasilocal energy is then defined in the $n$-dimensional surface 
$B\cap \Sigma_{\perp}$, a codimension
two submanifold,whose metric is
\be
ds^2=\frac{L^2}{\epsilon l^2}h_{ij}dx^i dx^j
\ee
and is given by the integral
\be
E(B\cap\Sigma_{\perp})=-\frac{L^{n-1}}{\kappa^2}\int_{B\cap\Sigma_{\perp}} \frac{1}{l^n\epsilon^{n/2}}\sqrt{h}d^n x 
(-n +\epsilon h^{kl}h^{ \prime}_{kl})
\ee
The divergences appearing in this expression have to be taken care of before
physical results can be obtained
(cf. \cite{Balasubramanian}\cite{Kraus}). We shall come back to this 
basic point in the next section.
\par

\section{The Conformal Anomaly}
The expression for the Brown-York quasilocal energy reads,
after expanding  $h_{ij}$ explicitly as a powers of $\e$, 
$h_{ij}=g_{ij}+\e h^{(1)}_{ij}+\e^2 h^{(2)}_{ij}+\mathcal{O}(\e^3)$
\be
E=\frac{1}{\kappa^{2}}    \int{d^{n}x\
  \frac{L^{n-1}}{l^{n}\ \e^{\frac{n}{2}}}\sqrt{|g|}\left[ n+\e \frac{n-2}{2} h^{(1)}
    +\e^2 \frac{n-4}{4}(2h^{(2)}-h^{(1)\ ij}\ {h^{(1)}}_{ij}+\frac{1}{2}{h^{(1)}}^{2}) + \mathcal{O}(\e^{3})\right]
  }
\ee
Expanding also the formulas in the Appendix (\ref{Lowenergy}) embodying 
 the Ricci-flatness condition on the ambient metric leads to
 the
relations
\bea
&&(n-2){h^{(1)}}_{ij}+h^{(1)}\ g_ {ij} - l^{2}R_{ij}(g)=0\nonumber\\ 
&&h^{(1)}=\frac{l^{2}}{2(n-1)}R\nonumber\\ 
&&h^{(2)}=\frac{1}{4}h^{(1)\ ij}\ {h^{(1)}}_{ij} 
\eea

and so on for higher dimensions.

For $d\in 2 \mathbb{Z}$ the term independent of $\e$ gives zero (in the form
($n-d$)) times the corresponding conformal anomaly. To be specific, in the two dimensional case,
\be
a_{(2)}= (n-2)\frac{L}{2 l^2{\kappa_4^2}}h^{(1)} = (n-2)\frac{L}{4{\kappa_4}^2}R\sim (n-2)E_2   \\
\ee
whereas in four dimensions
\bea
a_{(4)}= (n-4)\frac{L^3}{4 l^4 {\kappa_6}^2}(2h^{(2)}-h^{(1)\ ij}\
{h^{(1)}}_{ij}+\frac{1}{2}{h^{(1)}}^{2})=\nonumber \\
=\frac{-L^{3}}{32\kappa_6^2}(n-4)(R^{ij}\ R_{ij}-\frac{1}{3}R^{2})\sim (n-4)(E_4 + W_4)
\eea
where  $E_n$ is the integrand of the Euler character 
in dimension $d=n$ and
$W_n$ is the quadratic Weyl invariant.
 On the other hand, the quasilocal 
energy has to be refered to a particular template, which is to be attributed
 the zero of energy. In our case this would mean to substract the energy 
of the flat six dimensional space, and stay with
\be
E=- \frac{1}{\kappa^{2}}\int{    d^{n}x\ \frac{L^{n-1}}{l^{n}
\e^{\frac{n}{2}}}\sqrt{|h|}\e h^{ij}h_{ij}^{\prime}}
\ee
which is such that its finite part is 
 proportional to $E_4 + W_4$ with non-zero coefficient.
\par
It is indeed remarkable that this is the correct form (up to 
normalization) for the conformal anomaly for conformal invariant matter; 
this fact allows for an identification of the central function of the CFT, 
namely \footnote{Choosing $L=l$ in order not to introduce an extra arbitrary scale.},
\be
c \sim \frac{l^{4}}{\kappa_6^2}
\ee
It could be thought that the scale $l$ is arbitrary in our problem, because
the background is Ricci-flat; this is an illusion, however, because by 
dimensional analysis,  the Riemann squared scalar
(which determines , for example, the geodesic deviation equation) 
is proportional to 
$\frac{1}{l^4}$. 
\be
R^{\a\b\gamma\d}R_{\a\b\gamma\d}\sim\frac{1}{l^4}
\ee
We can then still refer to $l$ as the 
{\em radius of curvature} albeit in a generalized \footnote{
This argument fails for the flat background; then $l$ is really arbitrary.
}
sense. What is physically important is that string corrections are 
proportional to the curvature invariants, so that in order for them to be 
small $ l$ has to be large in string units: $l>>l_s$. 
\par
If we assume that (up to factors of order unity)
\be
\frac{1}{\kappa_6^2}=\frac{V_4}{g_s^2 l_s^8}
\ee
where $V_4$ is the volume of the compact manifold $C_4$,
and we assume that the boundary CFT is a gauge theory in the large $N$
limit, (so that the central charge scales as $c\sim N^2$) then this implies 
that
\be
l^4=\frac{g_s^2 l_s^8 N^2}{V_4}
\ee
so that $l\sim N^{1/2}$, which is different to AdS/CFT, in which 
$l\sim N^{1/4}$; the difference is  explained by dimensional
 analysis, owing to the fact that we now have to employ the six-dimensional 
Newton's constant instead of the five-dimensional one.

\section{Conclusions}
Some novel string backgrounds have been presented which seem to embody holographic
behavior, at least in the semiclassical regime. This behavior is of a different
kind from the one involved in the usual AdS/CFT duality in that there are two
holographic coordinates, of which one is timelike and the other spacelike.
This fact gives  the regulated five-dimensional boundary a dynamic 
character which we
have only partially explored. 
\par
In order for the system to retain some supersymmetry, then in the simplest 
case in which $A_6$ is flat the compact space $C_4$ must be a Calabi-Yau
twofold. The likely candidate for a ECFT is then a finite 
($\beta(g) = 0$) 
euclidean super Yang-Mills theory with $SU(N)$ gauge group,
 ${\cal N}=2$ supersymmetries and  broken $R$-symmetry. We indeed know how to
write down the central charge of this CFT on the gravitational side, namely
$ c=\frac{l^{4}}{\kappa_6^2}$. This would mean that $l\sim N^{1/2}$.
\par

The relationship of the backgrounds considered in 
this paper to the ones related to the Maldacena AdS/CFT conjecture (and, in particular,
the origin of the Ramond-Ramond fields from this point of view) still eludes
 us. Besides, the situation for the proposed duality has also some similarities with 
the dS/CFT duality of Strominger's (\cite{Strominger}), in the sense that here also, at least one of the
holographic coordinates is timelike.
\par
Further work is needed to clarify this relationship, as well as to
further expand the operator mapping.

\section*{Acknowledgments}
One of us (E.A.) acknowledges useful discussions with C\'esar G\'omez 
and Jose Barb\'on.
This work ~~has been partially supported by the
European Commission (HPRN-CT-200-00148) and CICYT (Spain).      


\appendix              
\section{$AdS_{(p,q)}$ and its horospheres}
Let us recall some elementary facts on horospheric coordinates (\cite{Alvarez}).
 For arbitrary $\pm$ signs, denoted by $\epsilon_{\m} = \pm 1$, the metric induced on the surface
\be\label{def}
\sum_{\m = 1}^{n+1}  \epsilon_{\m} x_{\m}^2 = \pm l^2
\ee
by the imbedding on the flat space with metric
\be
ds^2 =\sum_{\m = 1}^{n+1}  \epsilon_{\m} d x_{\m}^2 
\ee
can easily be reduced to a generalization of Poincar\'e's metric for the half-plane by introducing the coordinates
\bea
&&\frac{l}{z}\equiv x^{-}\nonumber\\
&&y^{i}\equiv z ~x^{i}
\eea
where we have chosen the two last coordinates, $x^{n}$  and $ x^{n+1}$ in such a way that their 
contribution to the metric
is $ dx_{n}^2 - dx_{n+1}^2$ (this is always possible if we have at least one timelike coordinate); and we define
$x^{-}\equiv x^{n+1} - x^{n}$. $ 1\leq i,j\ldots\leq  n-1$.
The generalization of the Poincar\'e metric is:
\be
ds^2 = \frac{\sum \epsilon_{i} dy_{i}^2 \mp l^2 dz^2}{z^2}
\ee
(where the signs are correlated with the ones defined in (\ref{def}), 
and the surfaces ${z=const}$ are called {\em horospheres} in the mathematical literature.
\par
The curvature scalar is given by:
\be
R=\pm \frac{n(n-1)}{l^2}
\ee
It is clear, on the other hand, that the isometry group of the corresponding manifold is
one of the real forms of the complex algebra $SO(n+1)$. The Killing vector fields are
explicitly given (no sum in the definition) by
\be
L_{\m\n}\equiv \epsilon_{\m}x^{\m}\pd_{\n}-\epsilon_{\n}x^{\n}\pd_{\m}
\ee
To be specific, when the metric is given by:
\be
ds^2=\frac{\delta_{ij}dx^i dx^j\mp l^2 dz^2}{z^2}
\ee
then the isometry group is $SO(n,1)$. This is the case for what could be 
 called 
{\em euclidean de Sitter}, $EdS_n$, which in our conventions has got 
all coordinates timelike, and {\em negative} curvature.
\par
The symmetric situation where
\be
ds^2=\frac{-\delta_{ij}dx^i dx^j\mp l^2 dz^2}{z^2}
\ee
enjoys $SO(1,n)$ as isometry group, and includes the ordinary de Sitter
space, $dS_n$. What one would want to call Euclidean anti de Sitter , 
$EAdS_n$, 
has got all its coordinates spacelike, and {\em positive} curvature.
\par
Finally, when the metric is given by
\be
ds^2=\frac{\eta_{ij}dx^i dx^j\mp l^2 dz^2}{z^2}
\ee
(where as usual, $\eta_{ij}\equiv diag(1,(-1)^{n-2})$),then the isometry 
group is $SO(2,n-1)$. This includes the regular Anti de Sitter, $AdS_n$.

\section{The Low Energy Limit}\label{Lowenergy}
If we characterize the metric in the $A_{n+2}$ ambient space as
\be
ds^2 = \frac{t^2}{l^2}h_{ij}(x,\rho)dx^i dx^j + \r dt^2 + t dt d\r
\ee\label{hs}
then its Ricci tensor reads
\bea\label{einstein}
&&l^2 R^A_{ij}=\rho [2 h''_{ij} - 2 h'_{il}h^{lm}h'_{mj} + 
h^{kl}h'_{lk} h'_{ij}]+ 
l^2 R_{ij}[h]-
(d-2)h'_{ij} - tr(h^{kl}h'_{kl})h_{ij}]\nonumber\\
&&=l^2 R_{ij}^{Bulk} -\frac{d}{\rho}h_{ij}\nonumber\\
&&R^A_{it}\equiv 0\nonumber\\
&&R^A_{i\rho}=\frac{1}{2}[h^{jl}(\nabla_j h'_{il} - \nabla_i h'_{jl})]
=R_{i\rho}^{Bulk}\nonumber\\
&&R^A_{\rho\rho}=-\frac{1}{2}[(h^{jk}h''_{kj}) -
\frac{1}{2} (h^{il}h'_{lm}h^{mn}h'_{ni})]=R_{\rho\rho}^{Bulk}+\frac{d}{4\rho^2}\nonumber\\
&&R^A_{\rho t}=0\nonumber\\
&&R^A_{tt}=0
\eea
where a prime means $\frac{d}{d\rho}$,and $\nabla_i$ is the covariant 
derivative
of the Levi-Civita connection of the metric $h_{ij}$. Demanding that this 
$(n+2)$-dimensional Ricci tensor vanishes reproduces 
the $(n+1)$-dimensional Einstein's equations
corresponding to a fixed cosmological constant
$\lambda=\frac{n(n-1)}{2 l^2}$ (cf. for example eq. (191) in \cite{Alvarez})

We have indeed represented by a superscript  the corresponding 
quantities in 
the five dimensional bulk space, which is defined as the manifold endowed with a metric:
\be
ds^2=- \frac{l^2}{4\rho^2}d\rho^2+\frac{1}{\rho} h_{ij}dx^i dx^j
\ee
Please refer to \cite{Alvarez} for an expansion in powers of $\e$ of this set of equations.

\section{The six-dimensional Ambient Space associated to flat four-dimensional space}
Let us work out the simplest possible example, namely, the $A_6$ ambient space 
corresponding to a flat four dimensional space. This  is, expressed in the canonical coordinates
introduced by Fefferman and Graham (FG), (\cite{fefferman})
\be
ds^2=-\frac{t^2}{l^2} d\vec{x}^2+\rho dt^2 + td\rho dt
\ee

In this case, there is a very simple scale and  invariance in the 
ambient space, namely
\bea
&&x\rightarrow\lambda x\nonumber\\
&&t\rightarrow \frac{1}{\lambda} t\nonumber\\
&&\rho\rightarrow \lambda^2 \rho
\eea

\par

Through the  study of  its geodesics, it is not difficult to find a change of 
coordinates which reduces it to flat six-dimensional space, namely,
\bea\label{change}
&& t\equiv \xi_0-\xi_5 \nonumber\\
&&x^i\equiv \frac{l \xi^i}{\xi_0-\xi_5}\nonumber\\
&&\rho\equiv \frac{(\xi^{\mu})^2}{(\xi_0-\xi_5)^2}\nonumber\\
\eea
where $(\xi^{\mu})^2=\xi_0^2-\xi_5^2-\vec{\xi}^2$. The flat space 
$\mathbb{R}^6$ with coordinates $\xi^{\m}\equiv (\xi^0,\vec{\xi},\xi^5)$, reads
\be
ds^2=\eta_{\m\n}d\xi^{\m}d\xi^{\n}
\ee
(where $\eta_{\m\n}=diag(1,(-1)^{5})$).
The inverse change of coordinates is given by:
\bea\label{inverse}
&&\xi^0=\frac{t}{2}(1+\rho+\frac{x^2}{l^2})\nonumber\\
&&\xi^5=\frac{t}{2}(- 1+\rho+\frac{x^2}{l^2})\nonumber\\
&&\xi^i=\frac{x^i}{l}t
\eea
Lorentz transformations enjoy a nonlinear realization in the FG coordinates.
The scale invariance in the FG coordinates, for example, 
 corresponds to a boost in the $\xi^5$ direction in the Minkowskian
 coordinates. If we perform a general Lorentz transformation 
$\xi^{\m}={\Lambda^{\m}}_{\n}\xi^{\n}=({\d^{\m}}_{\n}+{\theta^{\m}}_{\n})\xi^{\n}$
where
\be
\theta_{\m\n}= \left( \begin{array}{ccc}
0 & -\omega &  \frac{\vec{\a}+\vec{\b}}{2}  \\
\omega & 0 &   \frac{\vec{\a}-\vec{\b}}{2}  \\
-\frac{\vec{\a}+\vec{\b}}{2}  & -\frac{\vec{\a}-\vec{\b}}{2}  & 
\theta_{ij}\\
\end{array}    \right)
\ee
the corresponding change (linearized) in the FG coordinates reads
\be \begin{array}{l}
\d t= t (\omega+\frac{\vec{\a}\vec{x}}{l}) \\
\d \r= -2\r (\omega+\frac{\vec{\a}\vec{x}}{l})  \\
\d x^{i}= \frac{l}{2}b^i - \omega x^i +{\theta^{i}}_{j}x^{j} 
+\frac{l}{2}(\r +\frac{x^{2}}{l^{2}})a^i-\frac{\vec{\a}\vec{x}}{l}x^i
\end{array}
\ee

\par
The corresponding jacobian of the change of coordinates (\ref{change})
reads
\be
|\det{\frac{\pd \xi^{\m}}{\pd (\rho,y, x^i)}}|=\frac{1}{2 l^n}|t^{n+1}|
\ee
This means that 
there is a horizon at $t=0$, and we are only covering one-half of
Minkowski space, namely a {\em Minkowski wedge},
\be
\xi_0>\xi_{n+1}
\ee
There are then two interesting hypersurfaces in our problem:$B\equiv\{\rho=0\}$
(which we will call the {\em boundary}, a null surface, with null normal vector
$n^2=\frac{2\rho}{t^2}=0$), and $\Sigma\equiv\{t=0\}$ (which is the {\em 
horizon}
determining the portion of Minkowski space covered by the standard 
FG coordinates). The horizon is also a null surface, $n^2=0$.

Nothing prevents us from assuming that the coordinates $x^i$ live in a torus.
The relationship to the Minkowkski wedge is then lost.
For example, in the simplest case in which all coordinates live in a circle of radius
$L$, there is an equivalent  T-dual formulation (\cite{Giveon}) of the sort
\be
ds^2 = -\frac{l^2 }{t^2}d\vec{x}^2+\rho dt^2 + td\rho dt
\ee
and a dilaton
\be
\Phi=-\frac{n}{2}\log{\frac{t^2}{l^2}}
\ee
Non constant dilatons are notoriously difficult to work with; the original
 representation of the background will be then usually preferred.


\end{document}